\documentclass[letterpaper, 10 pt, conference]{ieeeconf}  

\IEEEoverridecommandlockouts                              

\usepackage{graphics} 
\usepackage{float}
\usepackage{epsfig} 
\usepackage{mathptmx} 
\usepackage{times} 
\usepackage{amsmath} 
\usepackage{amssymb}  
\usepackage{booktabs} 
\usepackage{multirow}
\usepackage{makecell}
\usepackage{url}
\usepackage{cite}
\usepackage[hidelinks]{hyperref}
\usepackage[capitalize]{cleveref}

\usepackage{_style}
\newcommand{\VarTime}{t}
\newcommand{\VarX}{x}
\newcommand{\SpikeTime}{t'}

\newcommand{\LifRate}{\rho}
\newcommand{\SoftLifRate}{\rho'}

\newcommand{\SynapticCurrent}{{I_s}}
\newcommand{\ThresholdCurrent}{{I_{th}}}

\newcommand{\MembraneCapacitance}{{C_m}}
\newcommand{\MembraneTimeConstant}{{\tau_m}}
\newcommand{\MembranePotential}{V}
\newcommand{\ThresholdPotential}{{\MembranePotential_{th}}}
\newcommand{\RestingPotential}{{\MembranePotential_0}}
\newcommand{\ResetPotential}{{\MembranePotential_{\text{reset}}}}

\newcommand{\RefractoryPeriod}{{\tau_{ref}}}
\newcommand{\SpikeRiseTime}{{\tau_{spike}}}

\newcommand{\SofteningParameter}{\gamma}
\newcommand{\SofteningNonlinearity}{\Theta}

\newcommand{\Parameters}{\theta}

\newcommand{\Action}{{a_{\VarTime}}}
\newcommand{\Observation}{{O_{\VarTime}}}

\newcommand{\Policy}{{\pi_{\Parameters}}}
\newcommand{\OldPolicy}{{\pi_{\Parameters}^{\text{old}}}}
\newcommand{\TrajectorySet}{{D_{\Parameters}}}
\newcommand{\Trajectory}{d}
\newcommand{\Advantage}{\hat{A}}

\newcommand{\PpoClipParam}{\epsilon}

\title{\LARGE \bf
S2Act: Simple Spiking Actor
}

\author{
Ugur Akcal$^{*135\dagger}$, Seung Hyun Kim$^{*2\dagger}$, Mikihisa Yuasa$^{1}$, Hamid Osooli$^{1}$, Jiarui Sun$^{1}$, Ribhav Sahu$^{3}$\\
Mattia Gazzola$^{267}$, Huy T.~Tran$^{1}$, and Girish Chowdhary$^{345\dagger}$%
\thanks{This work was supported in part by Navy \#N00014-19-1-2373, ONR \#N00014-20-1-2249, NSF Expeditions \#IIS–2123781, NSF EFRI \#2515342}
\thanks{University of Illinois at Urbana-Champaign: $^{1}$Aerospace Engineering, $^{2}$Mechanical Science and Engineering, $^{3}$Computer Science, $^{4}$Agricultural and Biological Engineering, $^{5}$Coordinated Science Laboratory, $^{6}$Carl R. Woese Institute for Genomic Biology $^{7}$National Center for Supercomputing Applications}%
\thanks{$*$These authors contributed equally to this work.}%
\thanks{$\dagger$Corresponding author: \{skim449,makcal2,girishc\}@illinois.edu}%
\thanks{This work has been submitted to the IEEE for possible publication. Copyright may be transferred without notice, after which this version may no longer be accessible.}%
}

\begin{document}

\maketitle
\thispagestyle{empty}
\pagestyle{empty}


\begin{abstract}
Spiking neural networks (SNNs) and biologically-inspired learning mechanisms are attractive in mobile robotics, where the size and performance of onboard neural network policies are constrained by power and computational budgets. Existing SNN approaches, such as population coding, reward modulation, and hybrid artificial neural network (ANN)–SNN architectures, have shown promising results; however, they face challenges in complex, highly stochastic environments due to SNN sensitivity to hyperparameters and inconsistent gradient signals. To address these challenges, we propose simple spiking actor (S2Act), a computationally lightweight framework that deploys an RL policy using an SNN in three steps: (1) architect an actor-critic model based on an approximated network of rate-based spiking neurons, (2) train the network with gradients using compatible activation functions, and (3) transfer the trained weights into physical parameters of rate-based leaky integrate-and-fire (LIF) neurons for inference and deployment. By globally shaping LIF neuron parameters such that their rate-based responses approximate ReLU activations, S2Act effectively mitigates the vanishing gradient problem, while pre-constraining LIF response curves reduces reliance on complex SNN-specific hyperparameter tuning. We demonstrate our method in two multi-agent stochastic environments (capture-the-flag and parking) that capture the complexity of multi-robot interactions, and deploy our trained policies on physical TurtleBot platforms using Intel’s Loihi neuromorphic hardware. Our experimental results show that S2Act outperforms relevant baselines in task performance and real-time inference in nearly all considered scenarios, highlighting its potential for rapid prototyping and efficient real-world deployment of SNN-based RL policies.
\end{abstract}

\section{Introduction}
\label{sec:intro}

Spiking neural network (SNN)-based reinforcement learning (RL) is a growing area of research at the intersection of neuroscience, machine learning, and robotics.
SNNs represent a broad class of biologically-inspired approaches for developing intelligent systems, alternative to traditional artificial neural networks (ANNs), that leverage continuous dynamics and event-driven mechanisms similar to biological neurons.
An important benefit of this event-driven framework is substantial energy savings in processing neural network models when used with neuromorphic hardware \cite{gruel2024neuromorphic,paredes2024fully,orchard2021efficient,davies2021advancing}.
Beyond energy efficiency, the sparse and asynchronous, event-driven nature of spike-based representations can also benefit multi-agent systems by reducing communication overhead and enabling decentralized, low-latency coordination.
Integrating SNNs with RL has therefore attracted considerable attention for mobile and autonomous robotics \cite{roy2019towards,rathi2023exploring}, to support more efficient on-board inference \cite{palossi2019nanodrone,tang2020reinforcement}, control \cite{tang2021deep,oikonomou2023hybrid}, and decision-making \cite{lobo2020spiking,eshraghian2023training}.
Realizing these benefits in complex, stochastic, real-world settings, however, remains a significant challenge.


\begin{figure}
    \centering
    \vspace{5pt}
    \includegraphics[width=\linewidth]{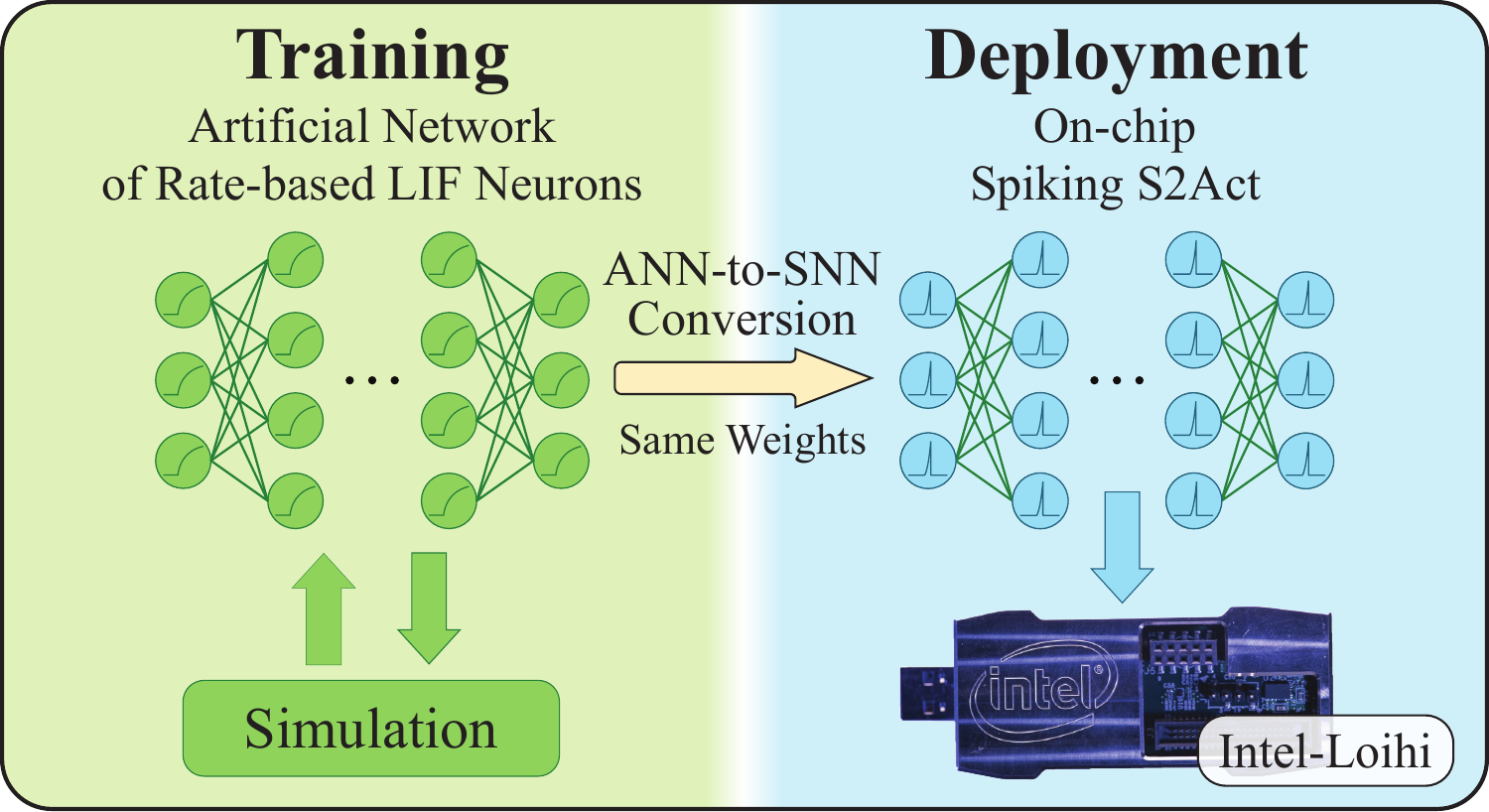}
    \vspace{-20pt}
    \caption{\textbf{S2Act.} An ANN with rate-based LIF activation is used for training, with learned weights then converted to a spiking network for deployment on neuromorphic hardware.}
    \vspace{-15pt}
    \label{fig1:overview}
\end{figure}

Existing approaches to SNN-based RL range from biologically-plausible learning rules to theory-driven methodologies; regardless of the paradigm though, current approaches share several common challenges.
For example, existing methods often require extensive tuning of neuronal hyperparameters due to the use of complex nonlinear neuron models \cite{zenke2021remarkable}, rely on large or complex network structures for auxiliary losses \cite{deng2023surrogate,cachi2023improving}, or suffer from vanishing gradients due to spiking dynamics that are incompatible with backpropagation \cite{Guo2024SpikeBP,lee2016training}.
The resulting instability and sensitivity of these methods are then often amplified in highly stochastic environments, such as multi-agent and adversarial settings found in real-world mobile robot deployments \cite{roy2019towards,zhang2021multi}.
Motivated by these challenges, we propose a simple yet effective ANN-to-SNN solution that matches or exceeds the performance of existing methods, while being computationally-efficient enough to enable rapid prototyping and swift sim-to-real deployment on mobile robots with neuromorphic hardware.
Our approach is illustrated in \Cref{fig1:overview}---our key idea is to configure a global set of neuronal parameters such that the resulting LIF response curves approximate the behavior of rectified linear unit (ReLU) activations \cite{hara2015analysis}, following principles from activation function design in deep learning \cite{glorot2011deep}.
This design choice removes the need to tune neuronal hyperparameters and facilitates gradient-based optimization while preserving the core advantages of SNNs.
This, in turn, enables stable and efficient offline training and further supports better sim-to-real deployment in real world environments.

\begin{figure*}[ht!]
    \centering
    \vspace{5pt}
    \includegraphics[width=\textwidth]{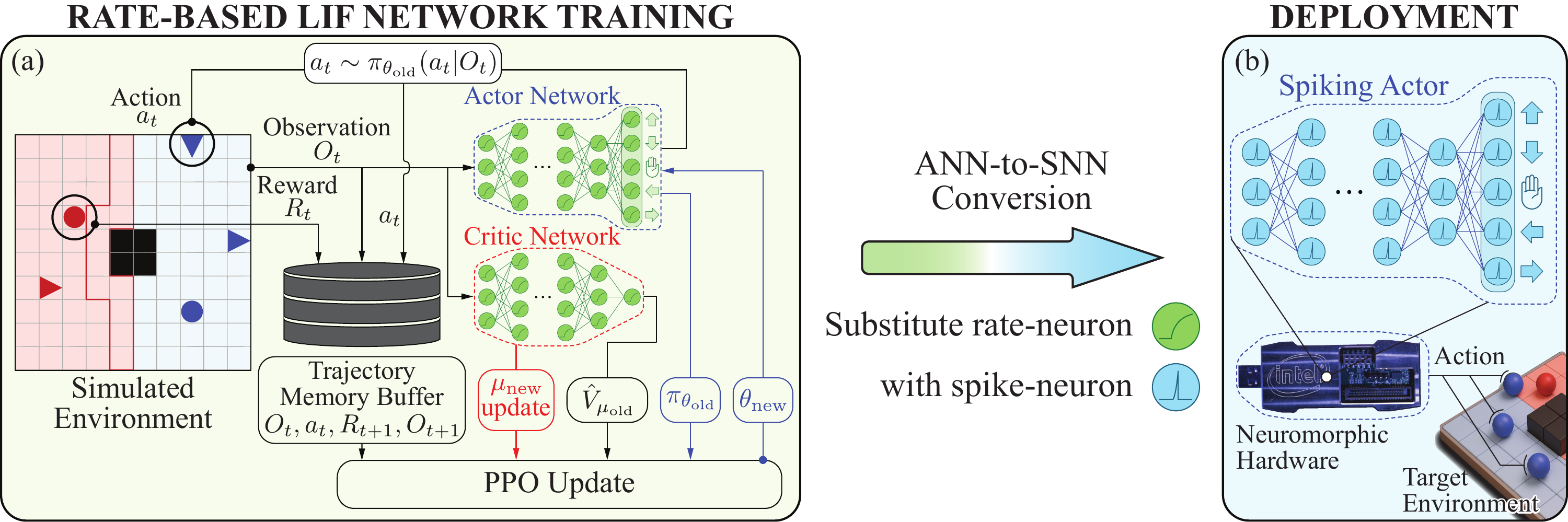}
    \vspace{-20pt}
    \caption{\textbf{S2Act training-to-deployment pipeline.} (a) We employ an ANN-to-SNN conversion strategy, enabling neuromorphic deployment of computationally lightweight SNN RL agents. In the rate-based LIF network training phase, the simulated environment provides discrete visual observations $\Observation$. These are passed to an actor-critic model, comprising a policy (actor) network and a value (critic) network, both modeled with soft-ReLLIF units (green circles).
        We use PPO to train these networks.
        (b) Once trained, the soft-ReLLIF neurons in the actor network are replaced with spiking LIF neurons (blue circles) for deployment on neuromorphic hardware, such as Intel’s Loihi.}
    \label{fig2:deploy_workflow}
    \vspace{-15pt}
\end{figure*}

The main contributions of this work are threefold: (1) we introduce a novel ANN-to-SNN conversion technique that mitigates the vanishing gradient problem in rate-based spiking models by shaping LIF neuron dynamics to approximate ReLU activations—enabling stable and gradient-friendly training; (2) we propose \textbf{S}imple \textbf{S}piking \textbf{Act}or (S2Act), a simple and computationally lightweight SNN actor for complex, stochastic environments; and (3) we validate S2Act in both simulated and real-world multi-robot settings, presenting the first on-chip demonstration of an SNN-based RL policy in a multi-agent adversarial task.

The rest of this paper is organized as follows: \Cref{sec:prev_work} discusses related work using SNNs in RL. \Cref{sec:method} details our proposed method. \Cref{sec:experiments} outlines our experimental setup and \Cref{sec:results} discusses our experimental results. Finally, \Cref{sec:conclusions} provides concluding thoughts, elaborates on the limitations of our method, and identifies potential directions for future work.

\section{Previous Work}
\label{sec:prev_work}


Recent research has explored a variety of techniques for employing SNNs within RL frameworks.
Many approaches emphasize biological plausibility by adopting learning rules inspired by neuroscience, such as R-STDP \cite{capone2024towards,lu2021autonomous,bellec2020solution,bing2019end}.
While these methods align with prevailing hypotheses about how biological neurons learn \cite{brzosko2019neuromodulation}, they often require extensive hyperparameter tuning and, in general settings, underperform compared to backpropagation \cite{deng2020rethinking}.
As a result, the training overhead limits rapid prototyping and swift sim-to-real deployment.
A second line of work combines SNNs with ANNs to leverage mature training pipelines while retaining spike-based computation.
For example, Chevtcenko and Ludermir \cite{chevtchenko2021combining} proposed an R-STDP tuned SNN integrated with a pre-trained binary convolutional neural network (CNN) for RL tasks.
Overall, these approaches are proposing the conceptual compatibility of SNN with ANN which our method is inspired from.
At the same time, many prior studies place less emphasis on neuromorphic hardware constraints, which often leads unstable solutions for deployment, leaving room to further improve implementation for real-world on-chip operation.


A third class of methods uses population coding \cite{auge2021survey,zhang2022multi,naya2021spiking}.
Tang et al. \cite{tang2021deep} proposed the population-coded spiking actor network (PopSAN), which integrates population coding with RL and includes many energy-efficient robotic control applications.
While inspirational and effective, population coding in SNNs comes with practical trade-offs, including layered architectural complexity from population redundancy \cite{pitkow2017inference}, increased resource demands \cite{pfeil2013six}, and additional training difficulties \cite{Neftci2019SurrogateLearning}.
In particular, population-coded architectures such as PopSAN typically require a large number of training parameters, since each state and action dimension is represented by a separate population of neurons.
In practice, these considerations may lead to larger and more entangled SNN architectures that require longer training times and are challenging to compile for neuromorphic chips \cite{rathi2023exploring}.

Collectively, prior SNN-based RL approaches have shown strong performance on relatively simple tasks but can face limitations in more complex and highly stochastic environments \cite{roy2019towards,zhang2021multi}.
Building on and motivated by these prior works, we aim to achieve more stable learning and reliable SNN policy performance, with a particular emphasis on handling highly stochastic environments, like multi-agent adversarial ones.



\section{Methodology}
\label{sec:method}

\begin{figure}
    \centering
    \vspace{5pt}
    \includegraphics[width=\linewidth]{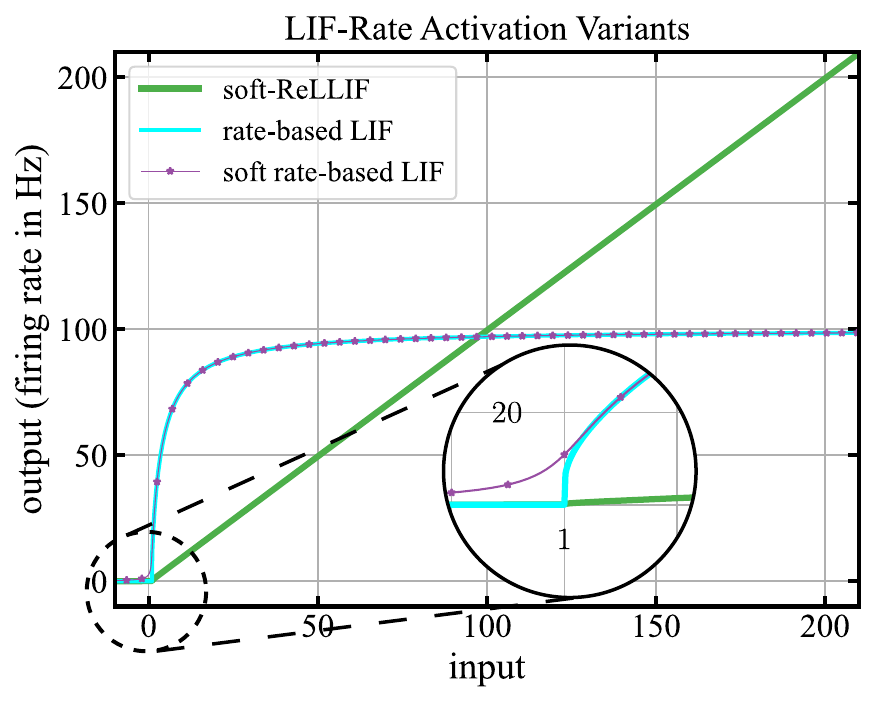}
    \vspace{-20pt}
    \caption{\textbf{Rate-based LIF neuron variants.} Rate-based LIF units (cyan line) exhibit unbounded gradients at the critical input value determined by their neuronal parameters. Although a smoothing term can be used to address these unbounded gradients, the resulting soft rate-based LIF unit (purple line with star markers) will still be prone to vanishing gradients. A common solution to this problem in networks with bounded activation functions is to employ ReLU activation. Therefore, we adjust the neuronal parameters of the soft rate-based LIF units to approximate the ReLU activation. Consequently, we utilize soft ReLLIF (green line) activations.}
    \label{fig3:RateLIF_variants}
    \vspace{-15pt}
\end{figure}

\Cref{fig2:deploy_workflow} illustrates the S2Act training-to-deployment pipeline, which follows the common paradigm of training offline and deploying online.
A rate-based actor-critic network is first trained using proximal policy optimization (PPO) \cite{schulman2017proximal} (see \Cref{sec:ann2snn}) in a simulated environment.
A key aspect of S2Act is that both the actor and critic use ReLU-like rate-based approximations of LIF neurons (soft-ReLLIF), whose parameters (i.e., neuron-to-neuron connection weights) are updated during training.
After training, the actor network is then converted to an equivalent spiking network (see \Cref{sec:s2act_architecture}), whose weights are used for deployment on neuromorphic hardware.

\subsection{ANN-to-SNN Conversion}
\label{sec:ann2snn}

The core idea underpinning our methodology is to design spiking neuron dynamics that closely emulate the behavior of ReLU activations, while carefully regularizing the dynamics such that they remain effectively expressible on neuromorphic hardware.
In practice, gradient methods for SNNs often struggle because neuromorphic systems rely on a narrow optimal range of spiking activity for reliable information representation.
This limitation stems from the nature of rate-based encoding: the resolution of neural activity (i.e., the rate of spiking events) depends on both inference time and activity magnitude.
The resulting narrow operating regime leaves little margin for shifts in weight distribution during training, which makes it more difficult to achieve reliable models under sensitive or noisy data collection.
By re-designing spiking neuron dynamics, maximally supporting the optimal range of neuron expression, we achieve a direct conversion paradigm with minimal discrepancy in sim-to-real settings.

The ANN-to-SNN conversion paradigm---i.e., training with differentiable activation units and substituting spiking neurons at inference---additionally avoids the complexities of surrogate design, spike encoding, and hyperparameter tuning.
This approach therefore maintains and utilizes the extensive capacity and scalable infrastructure of modern ANNs and backpropagation-based training, while keeping SNN dynamics optimal for neuromorphic deployment.
In this way, we bridge the gap between the efficiency of SNNs and the demands of training in complex, stochastic environments.

Our implementation involves few components beyond the soft-ReLLIF activation function: we regularize the activity range to support optimal rate-based value representation while controlling inference latency.
Unlike ANNs, where floating-point magnitude does not cost energy, activity regularization of spike-rate considerably affects both energy consumption and inference time, especially at the network scale.
To enforce this operating regime, we globally configure LIF neuron parameters such that their response curves approximate ReLU functions, while remaining compatible with gradient-based optimization, and rescale their activity to fit within the range of the planned deployment hardware.
This neuron-shaping approach addresses vanishing gradients and reduces the number of neuron parameters to tune, thereby easing stable training and deployment of deep SNN-based RL policies.


\begin{figure*}[t]
    \normalsize
    \centering
    \vspace{5pt}
    \includegraphics[width=\linewidth]
    {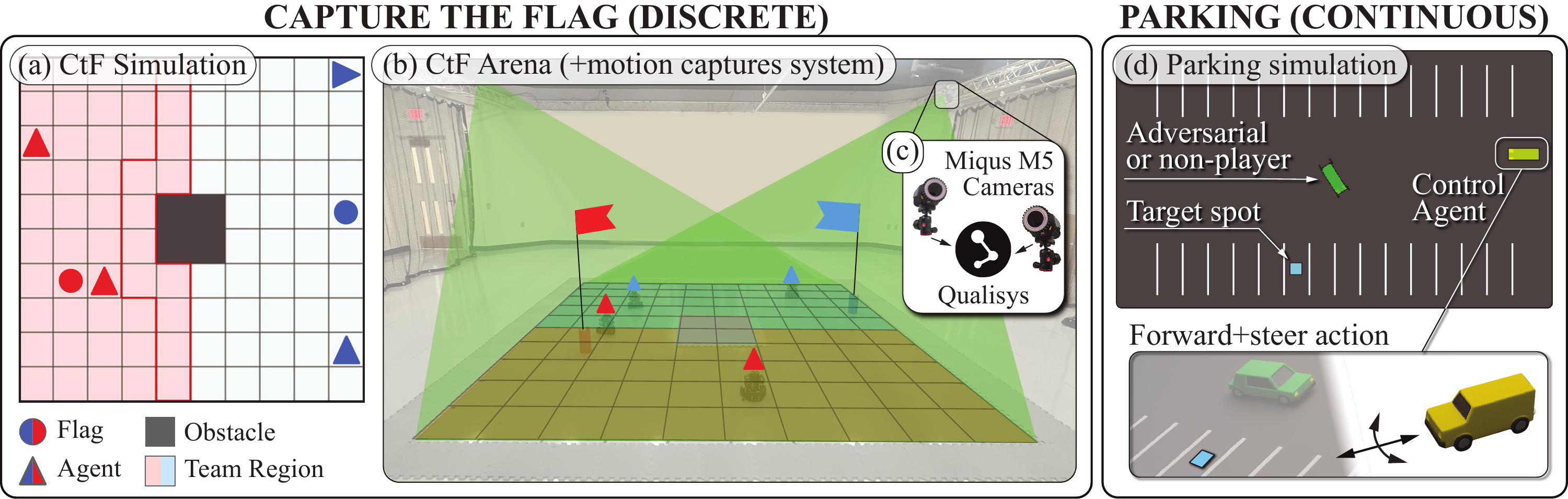}
    \vspace{-20pt}
    \caption{
        \textbf{Simulated and real-world environment for evaluations.}
        (a) Visualization of our simulated 2 vs. 2 CtF game. Gray squares are obstacles, triangles are agents, and circles are flags. The region highlighted by solid red lines is the border region for a red agent whose policy is the \textit{patrol} policy. The game has stochastic combat: when two agents are adjacent, whether one is eliminated depends on the territory, the number of nearby enemies, and the number of nearby allies.
        (b) Real-world CtF arena measures 12'\,$\times$\,12' and replicates the 10\,$\times$\,10 grid-world used during policy training. Blue, red, and gray tiles denote the blue team's territory, the red team's territory, and obstacles, respectively.
        (c) A Qualisys motion capture system with ten Miqus M5 cameras and Qualisys Track Manager software is used to track robot positions in real time.
        (d) Visualization of the parking environment. The green rectangle is the ego vehicle, and the blue square is the target parking spot.}
    \label{fig4:environments}
    \vspace{-15pt}
\end{figure*}

\subsection{SNN Implementation}
\label{sec:network-model}

The analytical firing rate solution $\LifRate[\SynapticCurrent(\VarTime)]$ of the sub-threshold dynamics of LIF neurons expressed by \Cref{eq:lif_rate} is the very core of the ANN-to-SNN conversion framework we implement.
\begin{align}
    \begin{split}
        \LifRate[\SynapticCurrent(\VarTime)] & =
        \left\{
        \begin{array}{cr}
            0                                                  & \text{if } \SynapticCurrent(\VarTime)\leq \ThresholdCurrent \\
            \left(\RefractoryPeriod+\SpikeRiseTime\right)^{-1} & \text{if } \SynapticCurrent(\VarTime)> \ThresholdCurrent
        \end{array}
        \right.                                                                                                                                                                                                                                                                                                  \\
        \ThresholdCurrent                    & = (\ThresholdPotential-\RestingPotential)\dfrac{\MembraneCapacitance}{\MembraneTimeConstant}                                                                                                                                                                      \\
        \SpikeRiseTime                       & = -\MembraneTimeConstant\log \left( 1-\dfrac{(\ThresholdPotential-\ResetPotential)\dfrac{\MembraneCapacitance}{\MembraneTimeConstant}}{(\RestingPotential-\ResetPotential)\dfrac{\MembraneCapacitance}{\MembraneTimeConstant}+\SynapticCurrent(\VarTime)} \right)
    \end{split}
    \label{eq:lif_rate}
\end{align}
\Cref{eq:lif_rate} gives the LIF firing rate $\LifRate$ as the time-averaged spike rate for a constant synaptic current $\SynapticCurrent(\VarTime)$.
At time $\VarTime$, $\SynapticCurrent (\VarTime)$ denotes the total synaptic current received by the neuron, $\MembraneCapacitance$ represents the membrane capacitance, $\MembraneTimeConstant$ is the membrane time constant, and $\RestingPotential$ is the resting potential of the neuron.
The membrane potential $\MembranePotential (\VarTime)$ functions as a scalar state variable that resets to a value $\ResetPotential$ whenever it reaches a threshold $\ThresholdPotential$.
The refractory period, denoted as $\RefractoryPeriod$, represents the duration a neuron needs before it can receive input currents again following a spike, where $\SpikeRiseTime$ specifies the time required for the neuron's potential to rise from $\ResetPotential$ to $\ThresholdPotential$ after a spike at time $\VarTime = \SpikeTime$, under the condition $\ThresholdPotential < \SynapticCurrent(\VarTime) = c \in \mathbb{R},\, \SpikeTime < \VarTime \leq \SpikeTime + \SpikeRiseTime$.
For a comprehensive discussion of the LIF neuron model and the analytical solution for its firing rate, we recommend consulting \cite{stockel2022harnessing}.

The rate-based LIF activation function characterized by \Cref{eq:lif_rate} is not compatible with backpropagation, since $\partial \LifRate / \partial \SynapticCurrent$ is discontinuous and unbounded at $\SynapticCurrent = (\ThresholdPotential - \RestingPotential)\MembraneCapacitance/\MembraneTimeConstant$. Nevertheless, a modification of \Cref{eq:lif_rate} following the approach of \cite{Hunsberger2015SpikingDNN} allows for a smoothed, rate-based approximation of the LIF model ($\SoftLifRate\left[ \SynapticCurrent(\VarTime) \right]$) as expressed by \Cref{eq:soft_lif_rate}.
\begin{align}
    \begin{split}
         & \SoftLifRate\left[ \SynapticCurrent(\VarTime) \right] = \left\{ \RefractoryPeriod+ \MembraneTimeConstant\log \left( 1+\dfrac{\ThresholdPotential}{\SofteningNonlinearity\left[ \SynapticCurrent(\VarTime) - \ThresholdPotential\right]} \right) \right\}^{-1} \\
         & \SofteningNonlinearity (\VarX) = \SofteningParameter \log\left( 1+e^{\VarX/\SofteningParameter} \right)
    \end{split}
    \label{eq:soft_lif_rate}
\end{align}

However, an activation function driven by \Cref{eq:soft_lif_rate} becomes vulnerable to vanishing gradients as its output spiking activity converges to the maximum firing rate. To overcome complications arising from output activity saturation, we draw inspiration from \cite{glorot2011deep} and adjust the neuronal parameters in \Cref{eq:soft_lif_rate} so that each rate-based LIF unit in the ANNs used for training approximates a ReLU activation function, as depicted in \Cref{fig3:RateLIF_variants}.
\subsection{Policy Optimization}

We train an actor network (i.e., policy network) $\Policy$ composed of soft-ReLLIF neurons with trainable parameters $\Parameters$, following a typical RL training procedures (see \Cref{fig2:deploy_workflow}a).
We specifically use PPO, which maximizes the clipped objective given in \Cref{eq:ppo_clip_objective}. 
\begin{align}
    \begin{split}
         & \Parameters_{\text{new}} = \underset{\Parameters}{\operatorname{arg\, max}}\, \dfrac{1}{|\TrajectorySet|T}\sum_{\Trajectory\in \TrajectorySet} \sum_{t=1}^{T} \min \left( f_{t},\, g_{t} \right)      \\
         & f_{t} = f\left( \OldPolicy, \Policy, \Observation, \Action \right) = \dfrac{\Policy (\Action|\Observation)}{\OldPolicy (\Action|\Observation)}\Advantage_{\OldPolicy}(\Observation,\Action)           \\
         & g_{t} = g\left( \OldPolicy, \Policy, \Observation, \Action \right) =                                                                                                                                  \\
         & \hspace{30pt} \text{clip}\left( \dfrac{\Policy (\Action|\Observation)}{\OldPolicy (\Action|\Observation)},\, 1-\PpoClipParam ,\, 1+\PpoClipParam \right)\Advantage_{\OldPolicy}(\Observation,\Action)
    \end{split}
    \label{eq:ppo_clip_objective}
\end{align}
Once training is completed, we substitute the actor neurons with spiking LIF neurons for inference.

\subsection{Architecture Details}
\label{sec:s2act_architecture}

We keep the size of S2Act minimal to simplify implementation on neuromorphic chips, which are often not well-optimized for spatially complex layer structures (e.g., CNNs).
The actor network comprises two dense layers of 64 neurons each, with inputs represented as a vector of the positions of the environment components.
For neuronal parameters, we set $\MembraneCapacitance = 1$, $\MembraneTimeConstant = 1$, $\RestingPotential = 0$, $\ResetPotential = 0$, $\RefractoryPeriod = 0$, and $\SofteningParameter = 0.003$.
We used PyTorch to build and train the SNNs.
\section{Experimental Setup}
\label{sec:experiments}

We demonstrate S2Act in a simulated capture-the-flag (CtF) game, a real-world multi-robot implementation of that CtF game, and a simulated parking environment, as illustrated in \Cref{fig4:environments}.
We compare against three baselines representing different classes of SNN-based RL methods: a population coding method (PopSAN) \cite{tang2021deep}, a hybrid method (Hybrid SNN) \cite{chevtchenko2021combining}, and a recurrent network method (RSNN) \cite{bellec2020solution}.
We conducted extensive parameter sweeps for each baseline and chose the hyperparameters that produced the best performance for each method.

\subsection{CtF Environment}
\label{sec:ctf_env}

Capture the Flag (CtF) is a multi-agent adversarial environment and a useful testbed for stress-testing RL training: it combines long horizons with sparse, delayed rewards, and it requires policies to respond to unpredictable opponent behavior.
Training in this setting is challenging because the episodic reward signal is often noisy and stochastic---conditions under which SNN-based RL frequently struggles to achieve the stability needed to learn strong policies.

We consider a simulated CtF game with $m$ allied agents (blue) versus $n$ enemy agents (red), based on \cite{yuasa2024generating} (\Cref{fig4:environments}a).
The environment has discrete state and action spaces, and its complexity and stochasticity arise from agent interactions, territorial advantages, and adversarial dynamics.
We adjust task difficulty by varying the enemy team size $n$ and by choosing the heuristic policy that the enemy agents follow: \textit{random-walk} or \textit{patrol}.
In \textit{random-walk}, enemies select actions uniformly at random, whereas \textit{patrol} biases movement toward the border region to defend the flag.
A team wins either by capturing the opponent's flag or by capturing all opponents.

Here, we report results for two settings: (1) 1v1 CtF against a \textit{random-walk} enemy, which is relatively easy to learn, and (2) 2v2 CtF against defensive \textit{patrol} enemies, where training often converges to conservative behavior, and stable learning can yield more aggressive strategies that successfully capture the flag.
Additional environment details will be provided in a public GitHub repository \cite{github-multigrid}.

\subsection{Multi-Robot CtF Sim-to-Real with Loihi}
\label{sec:real_CTF}

We also prepared a real-world demonstration of the CtF environment with robots using neuromorphic hardware (see \Cref{fig4:environments}b).
Our experimental setup includes multiple TurtleBot3 Burger robots in an indoor arena, where a ground station runs an on-chip S2Act centralized policy for the blue team, and autonomous heuristic policies for the red team.
Communication among system components was established over a local wireless network using a router and ROS 2 Humble~\cite{macenski2022ros2}, and a custom P-controller was used to drive robots with feedback from a Qualisys motion-capture system.

We deployed an S2Act centralized policy (trained in simulation) on a Loihi neuromorphic chip \cite{davies2021advancing}.
Loihi is a brain-inspired neuromorphic processor designed to run asynchronous spiking networks with low energy consumption.
We used the Kapoho Bay model for its modularity and accessibility, which allow us to assess performance and energy in mobile deployment settings.
We interfaced Loihi with NengoLoihi \cite{dewolf2020nengo} and NxSDK \cite{rueckauer2022nxtf} to convert the trained policy network into spiking neurons and synapses.
This ANN-to-SNN conversion re-compiles non-linear activation functions as neuron nodes and defines synaptic connections scaled by the learned weights.
Additional hardware tuning was required to rescale the activity range to the optimal expression range; details will be provided in a GitHub repository upon acceptance.

\subsection{Parking Environment}
\label{subse:park_env}


Finally, we include a simulated parking environment \cite{Moreira2021DeepRLParking} that evaluates SNN-based RL under physical dynamics, continuous state-action control, and adversarial disturbance (see \Cref{fig4:environments}d).
In this task, an ego vehicle (yellow) must navigate to a designated parking area while another vehicle (green) moves towards the same area.
Stochasticity comes from the other vehicle's randomly sampled acceleration and from the parking position.
An episode terminates when the ego vehicle successfully park, collides with the other vehicle, or crashes into a wall.
Additional environment details are provided in \cite{github-parking}.

\begin{figure*}[t]
  \centering
  \vspace{5pt}
  \includegraphics[width=\linewidth]{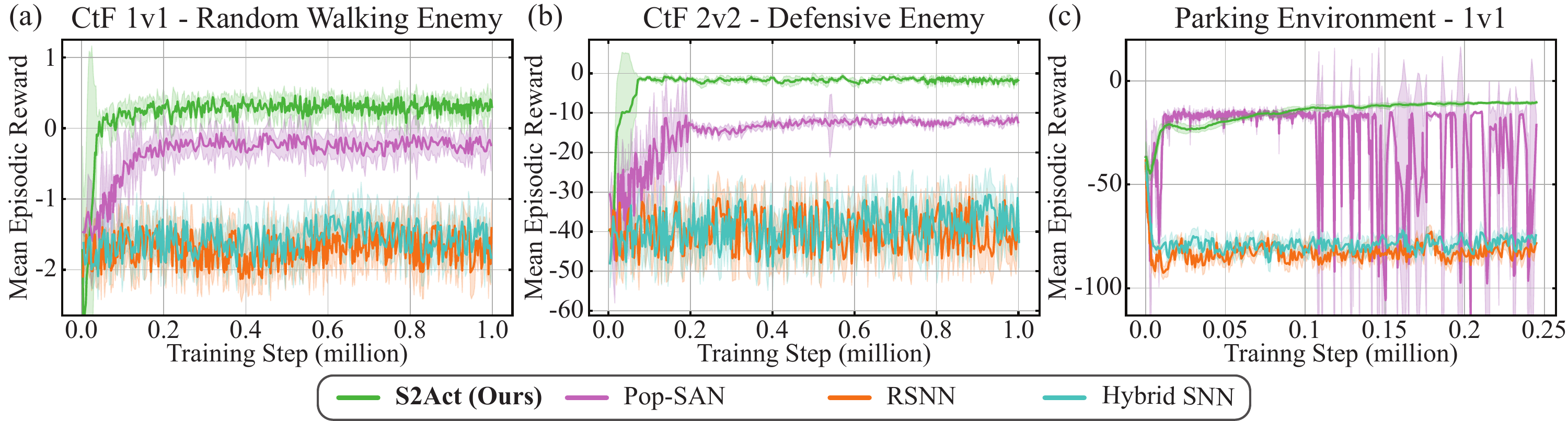}
  \vspace{-20pt}
  \caption{\textbf{Evaluation results.} Average training curves of S2Act and other baselines in two CtF scenarios (a, b) and the parking environment (c). Solid lines represent the average mean episodic reward over three seeds with shaded areas representing one standard deviation confidence intervals.
  }
  \label{fig5:rewards_plots}
  \vspace{-15pt}
\end{figure*}

\section{Results and discussions}
\label{sec:results}




\Cref{fig5:rewards_plots} shows training curves for our method, S2Act, and baselines in our simulated environments.
We see that S2Act consistently achieves the highest converged mean episodic reward and does so the fastest, with PopSAN emerging as the closest competitor.
We also observe that PopSAN's training becomes unstable after 100,000 time steps in the parking environment. Both RSNN and Hybrid SNN methods struggle to solve the tasks.

We also evaluated the performance of each fully trained agent over 1,000 episodes and report the mean values of various performance metrics in \Cref{tab:failure_analysis}.
Our metrics include the estimated \underline{e}nergy consumption \underline{p}er \underline{i}nference (EpI), the estimated \underline{m}ean \underline{in}ference \underline{t}imes (MInT), and the \underline{m}ean \underline{tr}aining \underline{t}ime (MTrT) for each policy.
EpI for Loihi was measured with a USB power meter to capture the collective power draw (including microprocessors and cooling), and EpI for CPU and GPU was estimated using the pyJoules package \cite{pyjoules}.
In the table, bold values mark the best EpI and success rates in each scenario.
For the CtF scenarios, we include statistics for the following success and failure cases: (1) Flag capture, where a controlled agent captures the enemy flag (success); (2) Defeated, where all controlled agents are captured by the enemy (failure); (3) Lost flag, where an enemy agent captures the friendly flag (failure); (4) Time-up, where the episode reaches the maximum step limit (failure).
For the parking environment, we include statistics for the following cases: (1) Success, where the car parks in the target spot; (2) Crash, where the car collides with the other vehicle. We tested both simulated and on-chip S2Act policies alongside our baselines.

The simulated S2Act agent achieved the highest success rate of $78.1\%$ and the fastest MTrT (2.4 hours compared to 8.3 hours by its closest competitor, PopSAN) by a significant margin in the 1v1 CtF random enemy scenario. The on-chip S2Act agent came second with a success rate of $72.6\%$. Similarly, both the simulated and on-chip S2Act policies delivered the best performance and fastest MTrT (5 hours versus 15.7 hours by PopSAN) in the 2v2 CtF defensive enemies scenario. It is worth noting that, in both scenarios, most failure cases occurred when the enemy had favorable initial conditions, either making it easier to defend its flag or capture the S2Act agents.
Furthermore, the low flag-capture rates of RSNN and Hybrid SNN are due to their inability to solve the task; episodes typically end in time-up as agents adopt conservative strategies rather than taking risk to capture the flag.
In the parking environment, the simulated S2Act also yielded the highest success rate and showed a significant improvement in MTrT relative to most baselines.
The on-chip S2Act policy performed slightly worse than PopSAN but outperformed the RSNN and Hybrid SNN baselines, which failed to solve the task.

\begin{figure}[t]
  \centering
  \vspace{5pt}
  \includegraphics[width=\linewidth]{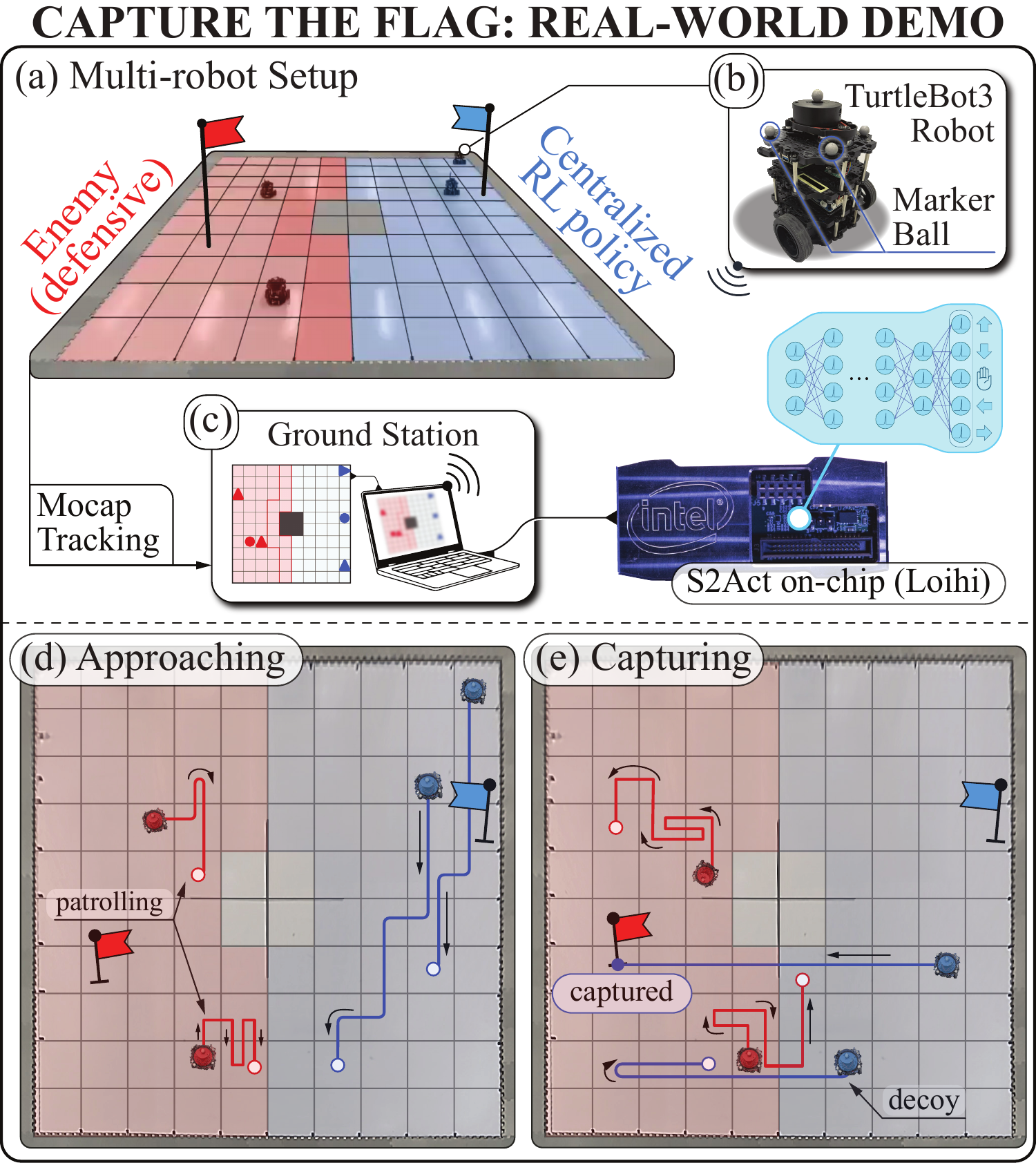}
  \vspace{-20pt}
  \caption{\textbf{Episode trajectory.}
    (a) Multi-robot demonstration of the on-chip deployment in the CtF task.
    (b) Each TurtleBot3 Burger robot is fitted with four sparsely placed marker balls to enable robust pose tracking. Communication among the components is handled via a local wireless network running ROS 2 Humble~\cite{macenski2022ros2}. The evaluations are conducted in a 2v2 scenario, where the blue agents engage with defensive red team behaviors.
    (c) A ground station communicates the on-chip S2Act centralized policy's commands to the blue team, while red team policies run individually on the ground station.
    (d-e) We overlay the trajectories executed by robots controlled by the S2Act policy network. A policy first moves agents to the nearest territorial boundary from the target flag, after which the agents coordinate to capture the flag.
    A supplementary video is provided.
  }
  \label{fig6:trajectory}
  \vspace{-15pt}
\end{figure}

We deployed the same on-chip S2Act policy in the real-world 2v2 CtF setup described in \Cref{sec:real_CTF}: TurtleBot3 Burger robots in an indoor arena, with the blue team controlled by the Loihi-based policy and the red team by heuristic policies, using motion capture and a ground station over a local network (see \Cref{fig6:trajectory}abc).
In repeated runs, the blue team successfully exhibited coordinated CtF behaviors---approaching the enemy flag, avoiding or engaging red agents, and capturing the flag when conditions allowed (see \Cref{fig6:trajectory}de)---without any fine-tuning in the physical environment.
We initially observed performance degradation due to out-of-range neural activity and a shift in the effective current range induced by thermal drift.
However, only minimal rescaling of neural activity was required to recover the same qualitative behavior observed in simulation, indicating robust sim-to-real transfer and low overhead for deployment on neuromorphic hardware.

Note that even with rescaling, we still observed a slight but noticeable reduction in the performance of the on-chip S2Act policy.
This gap is likely attributable to quantization effects on Loihi \cite{massa2020efficient}, arising from limited bit precision for representing synaptic weights and neural activations.
Although the policy’s qualitative behavior remains similar, residual discretization and timing effects can still reduce the control fidelity.
More broadly, the rate-based SNN implementation offers an inherent trade-off: achieving higher-precision encoding incurs latency, since each inference step consists of a temporal integration window to estimate output rates.
Consequently, rate-based encoding is fundamentally limited by the resolution of firing-rate estimates over finite windows: improving accuracy typically requires higher firing rates and/or longer windows, both of which increase latency or energy.
This limitation may be acceptable for higher-level planning and decision-making that is not time-critical, whereas fast continuous control will likely require alternative low-latency representations or decoding strategies.

\begin{table*}
  \vspace{5pt}
  \caption{\textbf{Performance measures.} Inference statistics and rate of successful termination.}

  \label{tab:failure_analysis}
  \centering
  \resizebox{\linewidth}{!}{
    \begin{tabular}{c|cc||ccc|cccc}
      \toprule
      \multirow{3}{*}{Environment} & \multirow{3}{*}{Method}  & \multirow{3}{*}{Hardware} & \multicolumn{3}{c|}{Inference statistics} & \multicolumn{4}{c}{Terminal conditions}                                                               \\
                                   &                          &                           & EpI                                       & MInT                                    & MTrT       & Flag Capture  & Defeated & Lost Flag & Time-up \\
                                   &                          &                           & [J / inf]                                 & [ms]                                    & [hours]    & (\%)          & (\%)     & (\%)      & (\%)    \\
      \midrule
      \multirow{5}{*}{1v1 CtF}
                                   & Simulated \textbf{S2Act} & GPU                       & 0.079                                     & 6.7                                     & 2.4        & \textbf{78.1} & 20.8     & 0.4       & 0.7     \\
                                   & \textbf{S2Act}           & Loihi                     & \textbf{0.033}                            & 27                                      & -          & \textbf{72.6} & 26.4     & 0.4       & 0.6     \\
                                   & PopSAN                   & GPU                       & 1.042                                     & 141.6                                   & 8.3        & 67.7          & 24.2     & 0.3       & 7.8     \\
                                   & RSNN                     & CPU                       & 0.468                                     & 57.8                                    & 12.1       & 4.2           & 17.3     & 0.1       & 78.4    \\
                                   & Hybrid SNN               & CPU                       & 0.066                                     & 9.2                                     & 27.3       & 0.3           & 17.1     & 0.8       & 81.8    \\
      \midrule
      \multirow{5}{*}{2v2 CtF}
                                   & Simulated \textbf{S2Act} & GPU                       & 0.093                                     & 7.2                                     & 5.0        & \textbf{34.0} & 65.0     & 0.0       & 1.0     \\
                                   & \textbf{S2Act}           & Loihi                     & \textbf{0.037}                            & 29                                      & -          & \textbf{29.4} & 61.6     & 0.0       & 9.0     \\
                                   & PopSAN                   & GPU                       & 1.415                                     & 188.1                                   & 15.7       & 21.2          & 62.5     & 0.0       & 16.3    \\
                                   & RSNN                     & CPU                       & 0.562                                     & 71.2                                    & 20.6       & 0.0           & 12.3     & 0.0       & 87.7    \\
                                   & Hybrid SNN               & CPU                       & 0.073                                     & 9.8                                     & 34.8       & 0.0           & 10.9     & 0.0       & 89.1    \\
      \toprule
      \multicolumn{3}{l||}{}       & EpI [J / inf]            & MInT [ms]                 & MTrT [hours]                              & Success Rate                            & Crash Rate                                                  \\
      \midrule
      \multirow{5}{*}{Parking}
                                   & Simulated \textbf{S2Act} & GPU                       & \textbf{0.012}                            & 1.4                                     & 4.9        & \textbf{43}   & 11.0                           \\
                                   & \textbf{S2Act}           & Loihi                     & 0.033                                     & 26                                      & -          & 39.1          & 14.4                           \\
                                   & PopSAN                   & GPU                       & 0.201                                     & 24.1                                    & 16.6       & \textbf{40.4} & 11.9                           \\
                                   & RSNN                     & CPU                       & 0.320                                     & 41.4                                    & 19.4       & 0.0           & 22.7                           \\
                                   & Hybrid SNN               & CPU                       & 0.03                                      & 2.3                                     & 34.9       & 0.0           & 21.4                           \\
      \bottomrule
    \end{tabular}
  }
  \vspace{-15pt}
\end{table*}

\section{Conclusions}
\label{sec:conclusions}

This paper presents S2Act, a computationally efficient RL architecture based on the ANN-to-SNN conversion paradigm. To the best of our knowledge, S2Act is the first on-chip SNN policy that uses an ANN-to-SNN conversion approach in an RL setting to tackle a real-world multi-agent adversarial scenario. Our experiments demonstrate that S2Act achieves superior training and inference performance in simulated CtF scenarios and parking tasks, outperforming other SNN competitors while using only two hidden layers of soft-ReLLIF neurons. Experiments further show that S2Act exhibits superior performance despite being significantly smaller in the number of trainable parameters and substantially more sample-efficient compared to its competitors, with minimal performance degradation due to ANN-to-SNN approximation errors.

These results underscore S2Act's robustness and adaptability in handling highly stochastic and adversarial environments, pushing the boundaries of current SNN capabilities. Hardware-in-the-loop simulations using Intel’s neuromorphic USB form factor, Kapoho Bay, demonstrate that on-chip S2Act policies maintain competitive, and in some cases superior, performance relative to their simulated counterparts, despite slight reductions due to hardware constraints.

Despite the promising results demonstrated by S2Act, several limitations warrant consideration. First, the reliance on rate-based training and subsequent ANN-to-SNN conversion introduces approximation errors that may degrade performance during on-chip inference, particularly under the quantization constraints of neuromorphic hardware like Loihi. While these discrepancies were modest in our experiments, they could become more pronounced in more complex tasks or larger networks. Additionally, S2Act currently assumes full observability of the environment and fixed neuronal parameters, limiting its adaptability to partially observable or dynamic contexts where real-time learning or plasticity may be beneficial.

Our findings highlight S2Act's potential for efficient, neuromorphic RL and real-world deployment of SNNs in complex, dynamic tasks. Future research may explore neuromorphic hardware-aware SNN architecture design to further optimize on-chip performance, as well as robust policy learning in adversarial and partially observable environments, broadening the practical applicability of SNNs in intelligent autonomous systems.




\bibliographystyle{ieeetr}
\bibliography{references,SHK}

\end{document}